\documentstyle[aps,prb,twocolumn,epsfig,floats]{revtex}
\begin{document} 
\draft
\twocolumn[\hsize\textwidth\columnwidth\hsize\csname @twocolumnfalse\endcsname
\title{
Secondary Coulomb Blockade Gap in a Four-Island Tunnel-Junction Array
}
\author{
Mincheol Shin,\cite{e-mail} Seongjae Lee, Kyoung Wan Park, and El-Hang
Lee}
\address{
Telecommunications Basic Research Laboratories\\
Electronics and Telecommunications Research Institute\\
Yusong POB 106, Taejon 305-600, Republic of Korea
}
\date{\today}
\maketitle

\begin{abstract}
In the ring-shaped tunnel-junction array with four islands, the
secondary Coulomb blockade gap in a low bias-voltage range is observed in
the $I-V$ characteristics. We attribute its appearance to the unique
topology of the array which induces up to two electrons to get trapped
inside. We have analyzed the formation and destruction of the gap in
terms of detailed single-electron tunneling processes. The negative
differential resistance behavior when the thermal and quantum
fluctuations are present is also studied.
\end{abstract}
\pacs{73.23.Hk, 73.23.-b}
]
\narrowtext 
\tighten

\section{Introduction}

\begin{figure}[bth]
\begin{center}
\leavevmode
\epsfig{width=1.0\linewidth,file=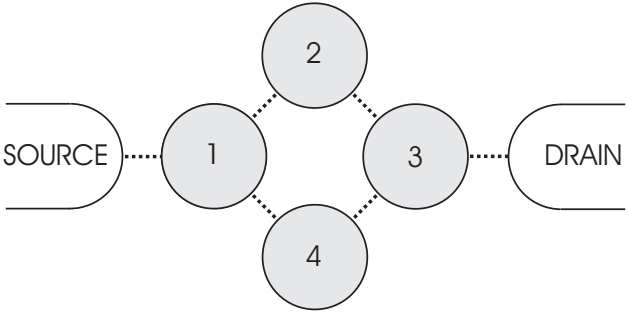}
\caption{
The four-island array. The tunnel-junctions are indicated by dotted
lines.
}
\label{fig:4dot-array}
\end{center}
\end{figure}

In a linear one-dimensional (1D) tunnel-junction arrays with identical
junctions,\cite{general,Amman,Hu,Bakhvalov,Whan,Likharev_Matsuoka}
electrons have no difficulty in traveling through the array once they
are injected into the array.\cite{Hu} In other words, electrons do not get
trapped inside the array, unless the characteristics of some junctions
of the array (e.\ g.\ junction resistance or capacitance) are
artificially altered so as to trap the electrons. On the other hand,
the ring-shaped array that is composed of two 1D arrays between the
source and the drain electrodes provides a natural way to trap the
electrons inside the array. In the four-island array of Fig.\
\ref{fig:4dot-array}, which has the smallest possible size for the
ring-shaped array, one electron may get trapped on island 1 (the
island that is closest to the source electrode) or two electrons on
islands 2 and 4 (the top and the bottom islands).\cite{my_jap} We have
recently reported that up to six electrons can be trapped inside the
sixty-island ring-shaped array.\cite{my_prl}

An important consequence of electrons' getting trapped inside the
ring-shaped array is appearance of the multiple Coulomb blockade (MCB)
gaps in the zero-temperature $I-V$ characteristics.\cite{my_prl} While
electrons are trapped inside, they block the electrical conduction
through the array so no current flows. Let us denote such zero-current
region with $n_t$ trapped electrons by ${\cal R}(n_t)$, and let us
suppose that up to $N_t$ electrons can be trapped inside. That
multiple electrons may get trapped implies that there exist multiple
regions ${\cal R}(1)$, ${\cal R}(2)$, $\cdots$, ${\cal R}(N_t)$ in the
$I-V$ curve where no current flows. If all of the insulating regions
are connected adjoiningly, a single gap in the form of an extended
Coulomb blockade (CB) gap will be seen in the $I-V$
characteristics. (Recall that a normal CB gap appears in the region
where there is no electron inside the array: i.\ e.\ ${\cal R}(0)$ by
our notation.) On the other hand, if neighboring regions ${\cal
R}(n_t)$ and ${\cal R}(n_t+1)$ are separated by a conducting region,
the MCB gaps will be seen in the $I-V$ characteristics. We have shown
that, in the ring-shaped array, during the transition from {\it odd}
to {\it even} $n_t$ there may exist such a conducting region, but
during the transition from {\it even} to {\it odd} $n_t$ no such
conducting region exists.\cite{my_prl} We qualitatively explained the
behavior by noting that the topology of the array demands that
dynamically unstable charge configurations must be passed during the
odd-to-even transition, but not necessarily during the even-to-odd
transition. Our argument leads us to state that the MCB gaps are a
unique transport property of the ring-shaped array, which cannot be
seen in linear 1D arrays, even with artificially trapped electrons.

In this paper, we will quantitatively describe the physics of the
single-electron tunneling in the ring-shaped array, by taking the
four-island array as the prototype system. Due to its simplicity, we
will be able to identify the detailed tunneling processes which leads
to the formation and destruction of the secondary Coulomb blockade
(SCB) gap (note that for the four-island array, only one extra CB gap
is possible, which we denote by the SCB gap.) Then we will be able to
discuss under which conditions the SCB gap appears, determine the
precise location of the gap, and predict the evolution of the gap as
the thermal and quantum fluctuations are introduced. Based on the
discussions, the physical properties of the MCB gaps of larger size
array may be similarly understood. We will also investigate the effect
of the uniform background charges on each island on the SCB gap.

\section{Theory}
\label{section:theory}


Our array system consists of four metallic or semiconducting islands,
positioned as depicted in Fig.\ \ref{fig:4dot-array}, between the
source and the drain electrodes. Nearest-neighboring islands are
connected by tunnel junctions so that there are two conducting paths
between the source and the drain: upper one via the top island (island
2) and lower one via the bottom island (island 4). Let us assign $C$
and $R$ for the tunnel-junction capacitance and resistance,
respectively, and $C_x$ for the cross-island capacitance (between
islands 1 and 3 and between islands 2 and 4). Furthermore, each island
is capacitatively coupled to the ground with same capacitance
$C_0$. The capacitance matrix $\bf A$ is then written as
\begin{equation}
{\bf A} = \left( 
\begin{array}{cccc}
D' & C & C_x & C \\
C & D & C & C_x \\
C_x & C & D' & C \\
C & C_x & C & D 
\end{array}
\right),
\label{eq:matrix_A}
\end{equation}
where $D' \equiv -3C-C_x-C_0$ and $D \equiv -2C-C_x-C_0$.

When a constant voltage $V$ is applied between the source and the
drain and $n_i e$ charges are there on island $i$ ($n_i$ is
an integer), the potential $\phi_i$ of island $i$ is given by, in the
matrix form,
\begin{equation}
\bf
\phi = A^{-1} Q,
\label{eq:A.phi=Q}
\end{equation}
where the column vector $\bf Q$ is defined by
\begin{equation}
\qquad\qquad\qquad Q_i = q_i - CV\delta_{i,1},\qquad\mbox{(i=1,\ldots,4)}\qquad
\label{eq:Q_i}
\end{equation}
where $q_i = n_i e$. By inverting the capacitance matrix given by Eq.\
(\ref{eq:matrix_A}), we have
\begin{equation}
\bf A^{-1} = \left({\begin{array}{cccc}
a_{+}&c\;&a_{-}&c\;\;\\
c\;\;&b_{+}&c\;\;&b_{-}\\
a_{-}&c\;\;&a_{+}&c\;\;\\
c\;\;&b_{-}&c\;\;&b_{+}
\end{array}}\right),
\end{equation}
where
\begin{eqnarray}
\nonumber
a_{\pm} &=&  \lambda_{+}/\tau_{-}+\lambda_{-}/\tau_{+}\pm\lambda_1/2, \\
b_{\pm} &=&  \lambda_{+}/\tau_{+}+\lambda_{-}/\tau_{-}\pm\lambda_2/2, \\
\nonumber
c &=& (\lambda_{-} - \lambda_{+})/2\sqrt{17},
\end{eqnarray}
where 
\begin{eqnarray}
\nonumber
\lambda_1^{-1} &=& D' - C_x, \\
\lambda_2^{-1} &=& D  - C_x, \\
\nonumber
\lambda_{\pm}^{-1} &=& D + C_x - (1\pm\sqrt{17})/2,
\end{eqnarray}
and
\begin{equation}
\tau_{\pm} \equiv (17\pm\sqrt{17})/2.
\label{eq:tau}
\end{equation}

The current $I$ is calculated for the constant applied voltage $V$ by
using the standard Monte Carlo (MC) method with the transition rates
determined by the Golden-rule;\cite{Amman,Bakhvalov} the transition
rate $\Gamma$ for a single-electron tunneling over a junction is given
by
\begin{equation}
\Gamma = {1 \over R e^2}{-\Delta F \over 1-\exp(\Delta F/k_B T)},
\label{eq:normal_rate}
\end{equation}
where $\Delta F$ is the free energy change by the tunneling event. The
free energy consists of the electrostatic energy contribution and the
work done by the external voltage source as usual. One may
conveniently obtain the free energy change $\Delta F$ by a tunneling
process involving islands $i$ and $j$ through the relationship\cite{Bakhvalov}
\begin{equation}
\Delta F = -{e\over 2}(\phi_i-\phi_j+\phi_i'-\phi_j'),
\label{eq:Delta_F}
\end{equation}
where $\phi_i$ and $\phi_i'$ are potentials of $i$th island before and
after the tunneling, respectively. Since only one electron is involved
in a single tunneling process, the charges on each island are changed by
$\pm e$ at most, so we can easily calculate $\Delta F$ via Eqs.\
(\ref{eq:A.phi=Q})-(\ref{eq:tau}) and Eq.\ (\ref{eq:Delta_F}).

We also used the master equation approach assisted by the MC method as
follows, especially when rare events should be taken care of. The
master equation\cite{Averin_Likharev,Fonseca} is written as
\begin{equation}
{dP_i \over dt} = \sum_{j=1}^{N^\ast} \Gamma_{ij}P_j - \sum_{j=1}^{N^\ast}
\Gamma_{ji} P_i
\label{eq:master_equation}
\end{equation}
where $\Gamma_{ij}$ is the transition rate from state $j$ to state $i$
and $P_i$ is the probability of finding state $i$ among total $N^\ast$
states considered in the calculation. The states considered in Eq.\
(\ref{eq:master_equation}) are composed of frequent states that are
readily obtained by a MC run and of rare states, if necessary, which
may have to be inserted manually with some physical reasoning if they
are hardly visited by the MC runs. Eq.\ (\ref{eq:master_equation}) is
solved either iteratively or by directly solving the equilibrium-state
equation:
\begin{equation}
\bf G P = 0
\end{equation}
where $\bf G$ is a $N^\ast$ by $N^\ast$ matrix where $G_{ij} =
\Gamma_{ij}-\delta_{ij}\sum_l \Gamma_{li}$ and $\bf P$ is the vector of
the state probabilities. Since the matrix $\bf G$ is singular, we take a
sub-matrix $\bf H$ of size $N^\ast-1$ by removing row $k$ and column
$k$ from the full matrix $\bf G$. Then
\begin{equation}
\left({\begin{array}{c}P_1 \\ \vdots \\ P_{k-1} \\ P_{k+1} \\ \vdots
\\P_{N^\ast}\end{array}}\right) = -P_k{\bf
H^{-1}}\left({\begin{array}{c}G_{1,k} \\ \vdots \\ G_{k-1,k} \\
G_{k+1,k} \\ \vdots \\ G_{N^\ast,k} \end{array}}\right),
\end{equation}
where $P_k$ is obtained by
\begin{equation}
P_k^{-1} = 1-\sum_{i,j=1}^{N^\ast-1} H^{-1}_{ij} G_{j+k,k}.
\end{equation}
In the master equation approach, {\it a priori} knowledge about the
relevant states is the key to the successful calculation.


We have also considered cotunneling\cite{cotunneling} up to the second
order. The reason that only up to the second-order is considered here
is that we will be interested in cotunneling out of the stationary
charge configuration, as will be discussed later, where the charges
are trapped on the top (island 2) and bottom (island 4) islands of the
array. The cotunneling rate is then given by the exact double-junction
formula:\cite{cotunneling}
\begin{eqnarray}
\nonumber
\gamma &=& {R_K \over 4\pi^2 e^2 R_1 R_2}\left\{2-\left(1+{2\over \Delta
F}{\Delta F_1\Delta F_2 \over \Delta F-\Delta F_1-\Delta F_2}\right)\right. \\
&\times&\left.\left(\sum_{i=1,2}\ln(1-\Delta F/\Delta F_i)\right)\right\}
\Delta F.
\label{eq:quantum_rate}
\end{eqnarray}
where $\Delta F_1$ and $\Delta F_2$ are the free energy changes for
the intermediate tunneling processes which are involved in the
cotunneling process, and $\Delta F$ is that for the cotunneling. In
the above equation, $R_K = h/e^2 \simeq 25.8 \rm{K}\Omega$, and $R_1$
and $R_2$ are the resistances of the junctions involved in the
cotunneling process.

In this paper, the units of the current, voltage, and temperature are
$\bar{I} = e/RC$, $\bar{V} = e/C$, and $\bar{T} = e^2/k_BC$,
respectively.

\section{Results And Discussions}
\label{section:results_and_discussion}

\begin{figure}
\begin{center}
\leavevmode
\epsfig{width=1.0\linewidth,file=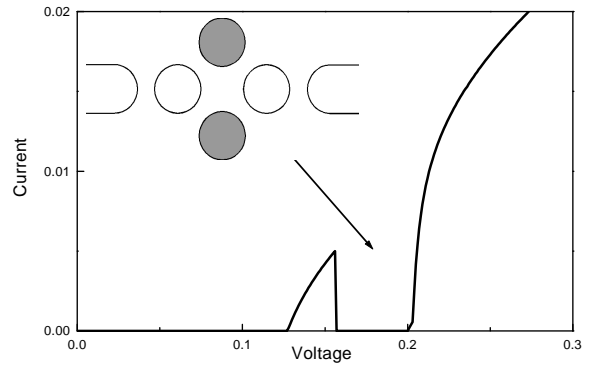}
\caption{ The $I-V$ characteristics of the four-island array for $C_0/C
= 3$ and $C_x/C = 0$ at zero temperature (thick solid line). In the
inset is shown the stationary charge configuration responsible for the
secondary Coulomb blockade gap in the interval [0.16,0.20].  The units
of the current and the voltage are $I/\bar{I}$ and $V/\bar{V}$,
respectively.}
\label{fig:iv-T0}
\end{center}
\end{figure}

A unique feature in the $I-V$ characteristics of the four-island
tunnel-junction array is the existence of another CB gap in a low
voltage region at low temperatures,\cite{my_jap} as shown in Fig.\
\ref{fig:iv-T0}. In the figure, where $C_0/C = 3$, $C_x/C = 0$, and
$T/\bar{T}=0$, the SCB gap in the interval of $[0.16,0.20]$ is seen in
addition to the primary CB gap in the interval of $[0,0.13]$. The
origin of the additional gap is attributed to the formation of the
trapped-electron configuration or the stationary charge configuration
(SCC) where two electrons trapped on the top and the bottom islands
block the electrical conduction,\cite{my_jap,my_prl} as schematically
shown in inset of Fig.\ \ref{fig:iv-T0}.

A SCC is a local minimum of the free energy in the configuration
space. That is, for each transition from the SCC to its adjacent
configurations (the configurations which can be reached by a
single-electron tunneling from the SCC) the free energy change $\Delta
F > 0$; otherwise, the electrons will not be trapped. In the following
we will describe the properties of the SCCs in the four-island array.

\subsection{Stationary Charge Configurations}

There are two SCCs that are noteworthy in the four-island array. The
first one is the SCC with one trapped electron on island 1 (SCC-1).
If we denote a charge configuration by $\{-n_i\}$ where $n_i$ is the
number of electrons on island $i$, SCC-1 corresponds to
$\{-1,0,0,0\}$. One of distinct features of the ring-shaped
four-island array in contrast to the linear 1D arrays is the existence
of the SCC-1. In linear 1D arrays of four islands with identical
junctions, the tunneling process which triggers the conduction is
$\{0,0,0,0\} \rightarrow \{-1,0,0,0\}$; once an electron tunnels onto
the first island connected to the source, it has no problem in
tunneling through the array. More specifically, let us denote $V_1$ as
the threshold voltage for entrance of one electron onto the array
($\{0,0,0,0\} \rightarrow \{-1,0,0,0\}$; let us denote the free energy
change of the process by $\Delta F_1$) and $V_t$ as the threshold
voltage for the electron to move forward ($\{-1,0,0,0\} \rightarrow
\{0,-1,0,0,\}$). In linear 1D arrays, $V_t < V_1$, which implies that
at the threshold voltage $V_1$, the free energy change $\Delta F_t =
F\{0,-1,0,0\} - F\{-1,0,0,0\}$ is already less than zero, so the
threshold voltage for electrical conduction is $V_1$ in that case. In
the ring-shaped four-island array, however, $V_1 < V_t$, as will be
seen shortly, which implies that there is a range $[V_1,V_t]$ where
the SCC-1 persists.\cite{Vt}

$\Delta F_1$ and $\Delta F_t$ of our four-island array are given by,
using Eqs.\ (\ref{eq:A.phi=Q})-(\ref{eq:tau}) and Eq.\
(\ref{eq:Delta_F}),
\begin{eqnarray}
\nonumber
\Delta F_1 &=& F\{-1,0,0,0\} - F\{0,0,0,0\} \\
           &=& -(1+a_{+})V - a_{+}/2 
\label{eq:F_1}
\end{eqnarray}
and
\begin{eqnarray}
\nonumber
\Delta F_t &=& F\{0,-1,0,0\} - F\{-1,0,0,0\} \\
           &=& -(c-a_{+})V - (b_{+}-a_{+})/2.
\label{eq:F_t}
\end{eqnarray}
The threshold voltages $V_1$ and $V_t$ are zeros of the above free
energy differences. For $C_x/C=0$ (the strong screening case), for
simplicity, we get from Eqs. (\ref{eq:F_1}) and (\ref{eq:F_t})
\begin{equation}
V_1 = { (C_0+1)(C_0+4) \over 2(C_0(C_0+3)(C_0+4)+2) }
\label{eq:V_1}
\end{equation}
and
\begin{equation}
V_t = { (C_0+1)(C_0+4) \over 2(C_0+2)(C_0^2+4C_0+1) }.
\label{eq:V_t}
\end{equation}
From Eqs.\ (\ref{eq:V_1}) and (\ref{eq:V_t}), one can easily see that
$V_t > V_1$ for all $C_0 > 0$. The same is true for arbitrary
$C_x/C_0$. We thereby show that SCC-1 exists in the bias-voltage range
of $[V_1,V_t]$ in our four-island array.

The second SCC, which is at the heart of our discussion here, is the
one with two trapped electrons on islands 2 and 4 (SCC-2), whose
charge configuration is $\{0,-1,0,-1\}$. To investigate the formation
and destruction of the SCC-2, we need to consider following two
tunneling processes: 1) the tunneling process $\{0,-1,0,0\}
\rightarrow \{-1,-1,0,0\}$ which leads to onset of the SCC-2 (once the
configuration $\{-1,-1,0,0\}$ is reached, the desired stationary
configuration will be eventually reached from there because the
tunneling process $\{-1,-1,0,0\} \rightarrow \{0,-1,0,-1\}$ is a
frequent process) and 2) the tunneling process $\{0,-1,0,-1\}
\rightarrow \{0,-1,-1,0\}$ (or, equivalently, $\{0,-1,0,-1\}
\rightarrow \{0,0,-1,-1\}$) which leads to break-up of the SCC-2. Let
us denote the free energy changes of the former and the latter
tunneling processes as $\Delta F_2$ and $\Delta F_b$,
respectively. Then,
\begin{eqnarray}
\nonumber
\Delta F_2 &=& F\{-1,-1,0,0\}-F\{0,-1,0,0\} \\
&=& -(1+a_{+})V - (a_{+}+2c)/2
\label{eq:DeltaF_2}
\end{eqnarray}
and
\begin{eqnarray}
\nonumber \Delta F_b &=& F\{0,-1,-1,0\} - F\{0,-1,0,-1\} \\
&=& (c-a_{-})V + ((a_{+}-b_{+}) + 2(c-b_{-}))/2.
\label{eq:DeltaF_b}
\end{eqnarray}

\begin{figure}
\begin{center}
\leavevmode
\epsfig{width=1.0\linewidth,file=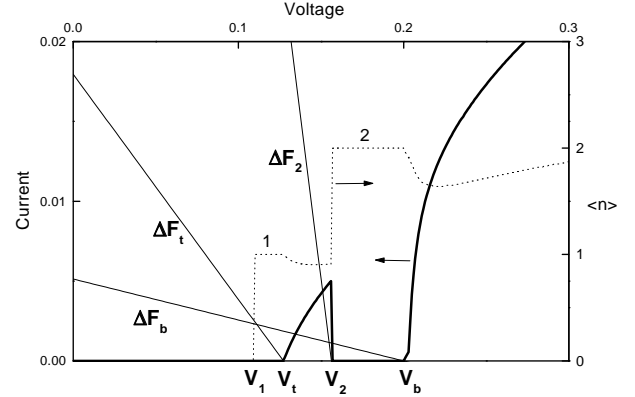}
\caption{The free energy differences $\Delta F_t$, $\Delta F_2$, and
$\Delta F_b$ (thin solid lines) are shown together with the $I-V$
curve of Fig.\ {\protect \ref{fig:iv-T0}} (thick solid line, left
axis). Also shown is the average number of electrons $\langle n
\rangle$ (dotted line, right axis). The threshold voltages $V_1$,
$V_t$, $V_2$, and $V_b$ are marked. The unit of the free energy
differences is $e^2/C$.}
\label{fig:iv-T0-freeE}
\end{center}
\end{figure}

At the threshold voltage $V_2$ where $\Delta F_2$ of Eq.\
(\ref{eq:DeltaF_2}) becomes zero, a second electron can now enter the
array, and after subsequent tunneling events, the charge configuration
$\{0,-1,0,-1\}$ will be eventually reached. From the configuration,
the only possible tunneling process is $\{0,-1,0,-1\} \rightarrow
\{0,-1,-1,0\}$ (or equivalently, $\{0,-1,0,-1\} \rightarrow
\{0,0,-1,-1\}$). If the free energy change for the process $\Delta
F_b$ is greater than zero, the tunneling process is energetically
unfavorable so that the electrons get trapped and consequently the
SCC-2 is established. See Fig.\ \ref{fig:iv-T0-freeE}. Thus the
necessary condition for the formation of the SCC-2 is $V_2 < V_b$,
where $V_b$ is the zero of $\Delta F_b$.

There are two tunneling processes which lead to destruction of the
SCC-2. The first one is the tunneling process where the trapped
electrons are forced to move forward: $\{0,-1,0,-1\} \rightarrow
\{0,-1,-1,0\}$ which is energetically favorable if $\Delta F_b \le 0$,
that is, if $V > V_b$. The other one is the tunneling process which
introduces a third electron into the array: $\{0,-1,0,-1\} \rightarrow
\{-1,-1,0,-1\}$, which is energetically possible only when $V > V_3$,
where $V_3$ is the threshold voltage for the process. Therefore, the
SCC-2 persists in the voltage range $[V_2,{\rm min}(V_b,V_3)]$.

Therefore, the SCC-2 is established if the following condition is
satisfied:
\begin{equation}
V_t < V_2 < {\rm min}(V_b,V_3).
\label{eq:blockade_condition}
\end{equation}
Then a current peak in the voltage interval of $[V_t,V_2]$ and the SCB
gap in the interval of $[V_2,{\rm min}(V_b,V_3)]$ will be seen in the
$I-V$ at zero temperature.

\begin{figure}
\begin{center}
\leavevmode
\epsfig{width=1.0\linewidth,file=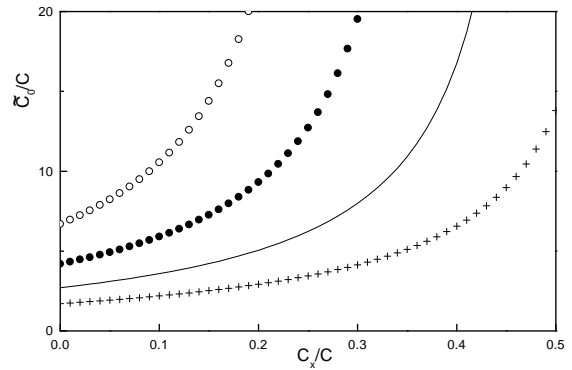}
\caption{The critical value $\tilde{C}_0/C$ for the observation of the
SCB gap at zero temperature with respect to the cross capacitance
$C_x/C$ for $q_0/e$ = 0.1 (crosses), 0 (solid line), -0.1 (solid
circles), and -0.2 (open circles). $q_0$ is the uniform background
charges induced on each island (see Section {\protect
\ref{section:ubc}}).}
\label{fig:critical-C0}
\end{center}
\end{figure}

Let us now discuss when the condition of Eq.\
(\ref{eq:blockade_condition}) holds. For $C_x/C = 0$, for
simplicity, threshold voltages $V_2$, $V_b$ and $V_3$ are given by
\begin{eqnarray}
\label{eq:V_2}
V_2 &=& {(C_0+3)(C_0+4)-2 \over 2(C_0(C_0+3)(C_0+4)+2) }, \\
\label{eq:V_b}
V_b &=& { C_0^2+C_0-4 \over 2(C_0+2)(C_0+1) },
\end{eqnarray}
and
\begin{equation}
V_3 \approx V_2 + (C/C_{eff})^2 e/C
\label{eq:V_3}
\end{equation}
where $C_{eff} = \sqrt{C_0^2+4C_0 C}$. From Eqs.\ (\ref{eq:V_t}) and
(\ref{eq:V_2})-(\ref{eq:V_3}), we obtain that the equality of Eq.\
(\ref{eq:blockade_condition}) is satisfied if $C_0 > \tilde{C}_0
\approx 2.7$. In general, when the cross capacitance $C_x$ is nonzero,
the expressions for the threshold voltages are much more complex than
in the strong screening case, and one has to resort to numerical
calculation for the evaluation of the equality of Eq.\
(\ref{eq:blockade_condition}). In Fig.\ \ref{fig:critical-C0}, we show
$\tilde{C}_0$, the critical value of $C_0$ for the observation of the
secondary Coulomb blockade gap, with respect to the screening factor
$C_x$. The figure shows that as the cross capacitance $C_x$ increases
$\tilde{C}_0$ also increases.

In summary, we have shown that the SCC-1 exists in the range ${\cal
R}(1) = [V_1,V_t]$ and the SCC-2 may exist in the range ${\cal R}(2) =
[V_2,{\rm min}(V_b,V_3)]$, if the condition $V_t < V_2 < {\rm
min}(V_b,V_3)$ is satisfied. The regions ${\cal R}(1)$ and ${\cal
R}(2)$ are zero-current regions, but in between the two regions, the
current flows such that it appears in the $I-V$ characteristics that
the SCB gap shows up in the range ${\cal R}(2)$. In terms of number of
trapped electrons $n_t$, we observe that the system becomes conducting
during the {\it transition} from $n_t$ = 1 to 2 (see Fig.\
\ref{fig:iv-T0-freeE} where the average number of electrons inside the
array $\langle n \rangle$ is also shown). Noting that both the SCC-1
and the SCC-2 are stable charge configurations in time, we remark that
the reason that the system becomes conducting during the transition is
that dynamically unstable charge configurations in time must be passed
{\it en route}. If we start with the SCC-1 and try to reach the SCC-2,
the charges should physically move in the meantime and they keep
tunneling through the array until the charges are balanced as required
by the geometry of the array (or electrostatic force). In short, the
SCB gap results because the topology of our ring-shaped array makes it
possible that {\it unstable conducting} state exists between two {\it
stable insulating} states. For the array of larger size having the
same topology (i.e. the array with two branches between the source and
the drain), this behavior is further generalized: that is, the current
peaks can appear during each transition from {\it odd} $\langle n
\rangle$ to {\it even} $\langle n \rangle$.\cite{my_prl} We have
reported in Ref.\ \onlinecite{my_prl} that three peaks can be observed
for the array of 60 islands, during each transition of $\langle n
\rangle$ = $1 \rightarrow 2$, $3 \rightarrow 4$, and $5 \rightarrow
6$. The reason that the current peaks arise only when $n_t$ changes
from odd to even is that the transition from even to odd $n_t$ does
not involve unstable charge states during the transition.

\subsection{Effect of Uniform Background Charges}
\label{section:ubc}

\begin{figure}
\begin{center}
\leavevmode
\epsfig{width=1.0\linewidth,file=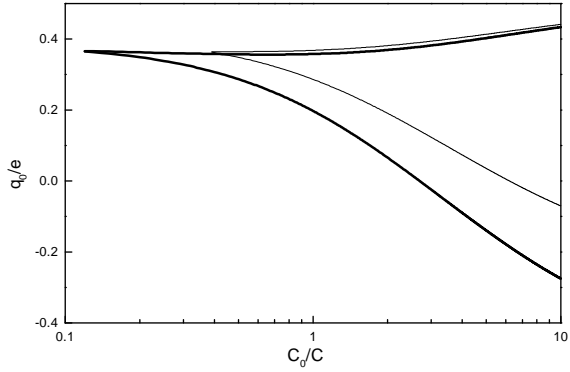}
\caption{The domain in the plane of $C_0/C$ and $q_0/e$ where the SCB
gap at zero temperature is observed, for $C_x/C = 0$ (the region
enclosed by thick solid lines) and $C_x/C = 1/4$ (the region enclosed by
thin solid lines).
}
\label{fig:mcg-domain}
\end{center}
\end{figure}

We have so far dealt with the system with neutral background charge
on each island: i.\ e.\ $q_i = n_i e$ in Eq.\ (\ref{eq:Q_i}). A
uniform background charge $q_0$ on each island may be induced by
attaching a (metallic) plate close to the array and applying a voltage
$V_g = q_0/C_0$ between the plate and the ground. Then $q_i = n_i e +
q_0$ in Eq.\ (\ref{eq:Q_i}),

The effect of the uniform background charge $q_0$ on the threshold
voltages introduced above is to shift them by a certain amount
linearly proportional to $q_0$ but at different rates for different
threshold voltages. For example, the threshold voltages $V_2$ and
$V_b$ become:
\begin{eqnarray}
V_2(q_0) &=& V_2(q_0=0) + q_0{a_{+}+a_{-}+2c \over 1+a_{+}} \\
V_b(q_0) &=& V_b(q_0=0) + q_0{b_{+}+b_{-}-a_{+}-a_{-}\over c-a_{-}}.
\end{eqnarray}
Therefore, if we vary $q_0$ in the range $-1/2 < q_0/e_0 < 1/2$, we
may tune it such that the equality in Eq.\
(\ref{eq:blockade_condition}) is satisfied. In Fig.\
\ref{fig:mcg-domain}, we show the domain in the plane of $C_0/C$ and
$q_0/e$ where the SCB gap can be observed at zero temperature for the
cases $C_x/C = 0$ and 1/4; the region enclosed by two thick solid
lines represents the domain where the SCB gap is observed at zero
temperature for $C_x/C = 0$, and that enclosed by two thin solid lines
for $C_x/C = 1/4$. Fig.\ \ref{fig:mcg-domain} indicates that for
$C_0/C \gtrsim 0.1$, the SCB region is readily accessible by adjusting
the (gate) plate voltage $V_g$. The region becomes wider as $C_0/C$
becomes larger, but if $C_0/C$ is too big ($> 10$) the SCB gap
shrinks, as does the usual Coulomb gap, and the current peak between
them becomes too narrow to be observed. We thus find that the range of
$0.1\lesssim C_0/C \lesssim 10$ is practically the range where the SCB
gap can be observed, by adjusting the gate voltage $V_g$. Fig.\
\ref{fig:mcg-domain} also shows that the effect of the screening
factor $C_x/C$ is to shrink the domain. That the domain shrinks with
respect to the increase of $C_x/C$ may be understood by noting that, if
the interaction between the top (2) and bottom (4) islands increases
by the increase of $C_x/C$, the stationary charge configurations are
harder to be built up, so one needs less coupling between islands, by
having bigger $C_0/C$, to compensate the increase in the interaction
between the islands.

We can infer from Fig.\ \ref{fig:mcg-domain} that, even when the
background charge $q_0/e$ of each island is induced, $\tilde{C}_0$
also increases as $C_x$ increases, as in the case of neutral
background charge (see Fig.\ \ref{fig:critical-C0}).


\subsection{Effect of Fluctuations}

We have so far investigated the SCB phenomena when the temperature is
zero and the quantum tunneling is suppressed. In the fluctuation-free
case, the existence of the SCB gap results from the fact that the SCC
with two trapped electrons is a stable configuration. The thermal
and/or quantum fluctuations will, however, destabilize the SCC, so
that the SCB gap will be transformed into a NDR region or entirely
disappear, depending on the degree of the fluctuations. We will first
discuss the effect of the temperature as follows.

\begin{figure}
\begin{center}
\leavevmode
\epsfig{width=1.0\linewidth,file=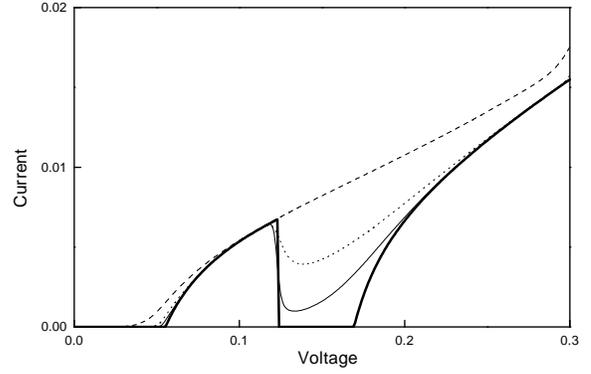}
\caption{The temperature evolution of the SCB gap for $C_0/C = 1.0$,
$C_x/C = 0.25$ and $q_0/e = 0.3$. The temperatures considered are
$T/\bar{T}$ = 0 (thick solid line), 0.0005 (thin solid line), 0.001
(dotted line), and 0.003 (dashed line).}
\label{fig:ivT}
\end{center}
\end{figure}

At finite temperature $T$, the trapped electrons in the SCC-2 have a
probability $\Gamma_{b}$ to tunnel forward onto the island 3:
\begin{equation}
\Gamma_b = {1\over R e^2}{-\Delta F_b \over 1 - \exp(\Delta
F_b/k_B T)}.
\label{eq:Gamma_th}
\end{equation}
The average time that it takes for the tunneling event to occur is
$\langle \Delta t_b \rangle = 1/2\Gamma_b$. (Tunneling out of both of
the islands 2 and 4 contributes to the factor of two.) That is, after
$\langle\Delta t_b\rangle$ on average, the SCC-2 is broken down by
tunneling of one of trapped electrons onto island 3. The electron
subsequently exits through the drain, provoking a series of tunneling
events to take place before the SCC-2 is restored, and the sequence of
break-up and restoration of the SCC-2 is repeated. Let us denote
$\langle\Delta t_a\rangle$ as the time that it takes for the SCC-2 to
be restored after it is broken down and $\langle\Delta Q_a\rangle$ as
the amount of charges that are transfered through the array
meanwhile. The current $I$ through the array is then given by
\begin{eqnarray}
\nonumber
I &=& \langle\Delta Q_a\rangle / (\langle\Delta t_a\rangle + \langle\Delta t_b\rangle) \\
  &=& I_a / (1 + 1/2\Gamma_{b}\langle\Delta t_a\rangle),
\label{eq:I_thermal}
\end{eqnarray}
where $I_a = \langle\Delta Q_a\rangle/\langle\Delta t_a\rangle$ is
approximately the current that would flow through the array if the
SCC-2 would not have been formed in the first place: i.\ e.\ the
current which would fill the SCB gap smoothly (see Fig.\
\ref{fig:ivT}). Since $\Gamma_{b} \sim \Delta F_b \exp(-\Delta
F_b/k_B T)$ at low temperatures, 
\begin{equation}
I \sim I_a \exp(-\Delta F_b /k_B T).
\end{equation}
Therefore, for $T < T_c \approx \Delta F_b / k_B$, the SCB gap at zero
temperature is transformed into a NDR region and for $T > T_c$, the
SCB gap entirely disappears and the current increases monotonically
with respect to the bias voltage. See Fig.\ \ref{fig:ivT}. We estimate
that $T_c/\bar{T} = k_B T_c/(e^2/C) \approx 10^{-3}$, which implies
that for the junction capacitance $C$ of the order of 1 aF, $T_c
\approx$ 1 K.

\begin{figure}
\begin{center}
\leavevmode
\epsfig{width=1.0\linewidth,file=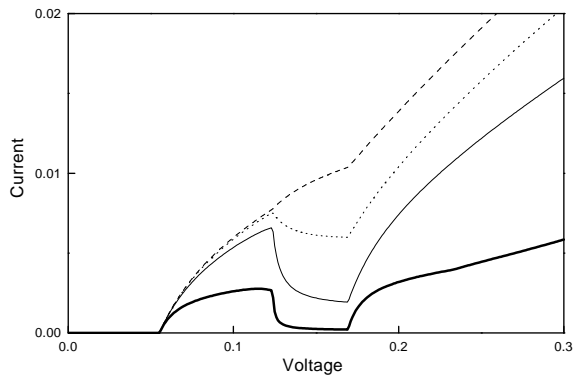}
\caption{The evolution of the SCB gap with respect to the cotunneling
strength at zero temperature. The array parameters are the same as in
{\protect Fig.\ \ref{fig:ivT}}. The resistances considered are $R_d/R$
= 10 (thick solid line), 1 (thin solid line), 0.25 (dotted line), and
0.1 (dashed line), where $R_d$ is the resistance of the junction
between island 3 and the drain electrode and $R = 1 {\rm M}\Omega$.}
\label{fig:ivQ}
\end{center}
\end{figure}

The second source of fluctuations is cotunneling. It is energetically
favorable that the trapped electrons in the SCC-2 cotunnel onto the
drain electrode. The cotunneling rate $\gamma_b$ is given by Eq.\
(\ref{eq:quantum_rate}) with $\Delta F_1 = \Delta F_b$, $\Delta F_2 =
a_{-}V -(a_{+}-4c)/2$ and $\Delta F = cV+(b_{+}+3b_{-})/2$. Once the
cotuneling event occurs, the same sequence of break-up and restoration
of the SCC-2 takes place as in the case of the thermal
fluctuation. The current $I$ is then given by Eq.\
(\ref{eq:I_thermal}) with $\Gamma_b$ replaced by $\gamma_b$. Since
$\gamma_b \propto R_K/R_d$, where $R_d$ is the resistance of the
junction between island 3 and the drain, one roughly has $I \propto
I_a/R_d$. Therefore, as in the case of the thermal fluctuation, there
exist a certain $R_d = R_c$ above which the SCB gap is transformed
into a NDR region and below which the $I-V$ characteristics show
monotonic increase. See Fig.\ \ref{fig:ivQ}. We estimate that $R_c
\approx 10^{5} \Omega$.

It may be worthwhile to discuss on how cotunneling was incorporated in
our calculation of the current in a little more detail. The current in
the SCB gap was obtained by solving the master equation with the
cotunneling process mentioned above included. As an alternative way,
we evaluated Eq.\ (\ref{eq:I_thermal}) with $\Gamma_b$ replaced by
$\gamma_b$ and by independently measuring $I_a$ through MC runs. Both
of the results agreed very well. Using the cotunneling rate $\gamma_b$
given by Eq.\ (\ref{eq:quantum_rate}) in the two methods poses no
difficulty in calculations in the SCB gap ($V_2 < V < V_b$). Beyond
the gap ($V \ge V_b$), however, $\gamma_b$ diverges because of the
occurrence of sequential tunnelings in parallel. Therefore we used the
approximation by Jensen and Martinis\cite{Jensen-Martinis} to
calculate the current beyond the gap and adjusted thus-obtained
current curve so as for the current to be continuous at the threshold
voltage $V_b$. This method of ours is far from being a rigorous one
but may be justified because we are only interested in the effect of
cotunneling in the SCB gap region for which we have treated the
problem more carefully.

\begin{figure}
\begin{center}
\leavevmode
\epsfig{width=1.0\linewidth,file=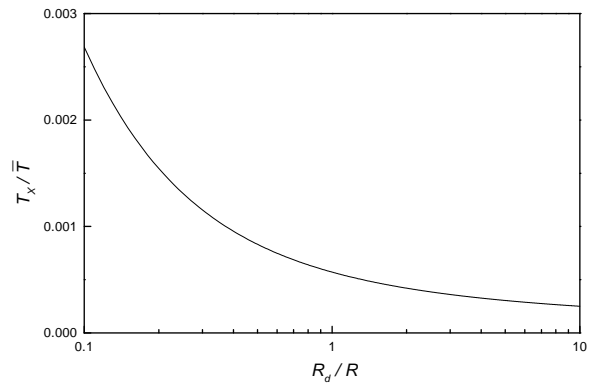}
\caption{The crossover temperature $T_x/\bar{T}$ with respect to $R_d/R$.
}
\label{fig:Tx}
\end{center}
\end{figure}

Discussing both the thermal and the quantum fluctuations separately,
we can now address the crossover temperature $T_x$ between the
quantum-fluctuation dominant regime and the thermal-fluctuation
dominant regime: for $T < T_x$ (the quantum regime), the currents
are virtually unchanged with respect to the temperature
but $T > T_x$ (the classical regime), the currents monotonically
increase with increase of temperature. We may estimate $T_x$ by
equating the normal transition rate $\Gamma_b$ and
the cotunneling rate $\gamma_b$ at a bias voltage
$V_m$ for which we take the middle point of the SCB gap. Fig.\
\ref{fig:Tx} shows thus-estimated crossover temperatures versus the
junction resistance $R_d$ for the case of $q_0/e = 0.3$, $C_x/C =
0.25$, and $C_0/C = 1.0$. As is expected, as the junction resistance
$R_d$ increases, the crossover temperature $T_x$ decreases in the
figure, which implies that the quantum fluctuations become less
effective as $R_d$ increases.

\subsection{Effect of Disorders}

Let us now discuss how rigid the SCB gap is against possible
imperfection of the array. Mainly there are three sources of disorder
in real experiments: 1) the size of the islands may differ from one
another, 2) the positions of the islands could be somewhat away from
the desired ones, and 3) the stray charges may be induced on each
island. Before we remark on the effect of the disorders in general,
let us take a specific example to demonstrate the effect of the stray
charges, among others. Note that stray charges on islands 2 and 4 may
substantially affect the formation of the SCC-2 because the charges
should be trapped there. When the background charge of island 2
differs from that of other islands by $\delta q$, the
threshold voltage $V_b$ changes by
\begin{equation}
\label{eq:delV_Vb}
\delta V_b = \left\{
\begin{array}{c} 
-\delta q {b_- - c \over a_- - c}\qquad {\rm if} \quad\delta q \ge 0 \\
-\delta q {b_+ - c \over a_- - c}\qquad {\rm if} \quad\delta q < 0,
\end{array}\right.
\end{equation}
and, regardless of the sign of $\delta q$, $V_2$ changes by
\begin{equation}
\delta V_2 = \delta q {c \over 1+a_+}.
\label{eq:delV_V2}
\end{equation}
The change in the threshold voltage $V_t$ is immaterial in the
discussion here. Eqs.\ (\ref{eq:delV_Vb})-(\ref{eq:delV_V2}) show that
the SCB gap shrinks linearly with respect to $\delta q$. The SCB gap
will be still seen if the change $\delta V_b - \delta V_2 < \Delta
V_{SCB} \equiv V_b - V_2$, and we find that the range in $\delta q/e$
for which the SCB gap persists is typically around 0.1. But the SCB
gap is much more vulnerable to negative stray charges than positive
ones, which reflects the fact that a negatively charged electron is
harder to get trapped if the island is more negatively charged. If we
can adjust the overall background charges on each island, in the
direction to compensate the stray charge, we may be able to re-enter
the region where the SCB gap is seen.
 
We found it a challenging task to treat the problem
analytically when the disorders are present in general, so we resorted to
numerical simulations with the following simple model. The first two
sources of imperfection mentioned above may be treated by relaxing the
condition of having identical islands and allowing the mutual
capacitances to have `random' contributions to some degree as follows:
\begin{equation}
\label{eq:C_ij_imperfect}
C_{ij} \rightarrow C_{ij}(1+\alpha \zeta_{ij}) \qquad {\rm for} \qquad i \neq j,
\end{equation}
where $C_{ij}$ refers to the capacitance between island $i$ and $j$
(including the source, the ground, and the drain electrodes), $\alpha$
is a constant adjusting the magnitude of the randomness and $\zeta$'s
are random numbers between -1/2 and 1/2. Note that in the above
equations, we have specifically put the subscripts for $\zeta$ to note
that the random numbers are differently assigned for different islands
and different pairs of islands. Similarly, for the third source of
imperfection, we may introduce random contributions to $q_0$ of each
island. Our numerical simulations show that for up to 10 \% of
random contributions the SCB phenomena is still observed, although the
SCB gap position or the NDR region is shifted and/or its width is
changed.

\section{Conclusion}
\label{section:conclusion}

In conclusion, we have shown that the secondary Coulomb blockade gap
of the four-island array at zero temperature is originated from the
property of the array that it traps up to two electrons inside, due to
its unique topology. The detailed tunneling mechanism which leads to
the formation and destruction of the stationary charge
configuration with two trapped electrons inside, which corresponds to
the additional gap, has been analyzed. We have also considered the
effect of the thermal and quantum fluctuations on the gap, and found
that the gap transforms into a NDR region upon introduction of the
fluctuations. The NDR feature suggests a potential application of the
ring-shaped array as a diode showing the NDR behavior. We have also
studied the effect of the uniform background charges on each island,
which can be realized in real experiments by attaching a ground plate
near the array and applying a voltage, and found that the secondary
Coulomb blockade gap is more readily accessible by adjusting the
voltage. Based on the detailed analysis that we have carried
out in this study, the physical properties of the multiple Coulomb
gaps of larger size array may be similarly understood.


The authors thank Gwang-Hee Kim for fruitful discussions on the
cotunneling effect and the crossover temperature. This work has been
supported by the Ministry of Information and Communications of Korea.


\begin{references}
\bibitem{Vt} 

In general, the tunneling process which leads to the conduction of the
four-island array is not necessarily the one mentioned in the text. For lower
values of $C_0/C$ ($C_0/C \lesssim 0.1$), for instance, two electrons
get trapped on islands 1 and 3 initially, and the tunneling process
which trigers the conduction is either the one where the trapped
electron on island 3 exits through the drain or the one where the
trapped electron on island 1 advances forward. When the uniform
background charges are induced on each island, as will be discussed
later, the situation becomes very complicated. However, for the range
of $C_0/C$ that we are interested in in this paper, the tunneling
process mentioned in the text is the one responsible for the
conduction.
\end{references}
\end{document}